\def\psra{PSR J0737--3039A}

\def\psrab{PSR J0737--3039A/B}

\def\nk{n_{\rm b}}

\def\Pb{P_{\rm b}}

\def\rfr#1{Equation\,(\ref{#1})}
\def\rfrs#1#2{Equations\,(\ref{#1})-(\ref{#2})}
\def\Rfr#1{Equation\,(\ref{#1})}
\def\Rfrs#1#2{Equations\,(\ref{#1})-(\ref{#2})}

\def\virg#1{``#1"}

\def\eqi{\begin{equation}}
\def\eqf{\end{equation}}
\def\eqia{\begin{eqnarray}}
\def\eqfa{\end{eqnarray}}

\def\rp#1#2{{#1\over#2}}
\def\lb#1{\label{#1}}

\def\bds#1{\boldsymbol{#1}}

\def\ton#1{\left(#1\right)}
\def\qua#1{\left[#1\right]}
\def\grf#1{\left\{#1\right\}}
\def\ang#1{\left\langle #1\right\rangle}
\documentclass[linenumbers]{aastex}

\usepackage{morefloats}
\usepackage[title]{appendix}
\usepackage{hyperref,textcomp}
\usepackage{booktabs}
\usepackage[table,xcdraw]{xcolor}
\usepackage{multirow}
\usepackage{rotating,tabularx}
\usepackage{float}
\usepackage{enumerate}
\usepackage{rotating}
\usepackage[polutonikogreek,english]{babel}
\usepackage{amsmath,starfont,textgreek,w-greek,wasysym}
\usepackage[flushleft]{threeparttable}
\usepackage{amsthm}
\usepackage{amscd,lineno}
\usepackage{amssymb,dsfont}
\usepackage{graphicx,epsfig}
\usepackage[utf8]{inputenc}
\usepackage[T1]{fontenc}
\usepackage{txfonts}
\usepackage{xr-hyper}

\RequirePackage{color}

\newcommand{\emaila}{lorenzo.iorio@libero.it}

\allowdisplaybreaks[1]

\begin{document}

\title{On the 2PN periastron precession of the Double Pulsar \psrab}

\shortauthors{L. Iorio}

\author{Lorenzo Iorio\altaffilmark{1} }
\affil{Ministero dell'Istruzione, dell'Universit\`{a} e della Ricerca
(M.I.U.R.)
\\ Viale Unit\`{a} di Italia 68, I-70125, Bari (BA),
Italy}

\email{\emaila}

\begin{abstract}
One of the post--Keplerian (PK) parameters determined in timing analyses of several binary pulsars is the fractional periastron advance per orbit $k^\mathrm{PK}$. Along with other PK parameters, it is used in testing general relativity once it is translated into the periastron precession $\dot\omega^\mathrm{PK}$. It was recently remarked that the periastron $\omega$ of \psrab\, may be used to measure/constrain the moment of inertia of A through the extraction of the general relativistic Lense--Thirring precession $\dot\omega^\mathrm{LT,\,A}\simeq -0.00060^\circ\,\mathrm{yr}^{-1}$  from the experimentally determined periastron rate $\dot\omega_\mathrm{obs}$ provided that the other post--Newtonian (PN) contributions to $\dot\omega_\mathrm{exp}$ can be accurately modeled. Among them, the 2PN one seems to be of the same order of magnitude of $\dot\omega^\mathrm{LT,\,A}$. An analytical expression of the total 2PN periastron precession $\dot\omega^\mathrm{2PN}$ in terms of the osculating Keplerian orbital elements, valid not only for binary pulsars, is provided elucidating the subtleties implied in correctly calculating it from  $k^\mathrm{1PN}+k^\mathrm{2PN}$  and correcting some past errors by the present author. The formula for $\dot\omega^\mathrm{2PN}$ is demonstrated to be equivalent to that obtainable from  $k^\mathrm{1PN}+k^\mathrm{2PN}$ by Damour and Sch{\"a}fer expressed in the Damour-Deruelle (DD) parameterization. $\dot\omega^\mathrm{2PN}$ actually depends on the initial orbital phase, hidden in the DD picture, so that $-0.00080^\circ\,\mathrm{yr}^{-1} \leq\dot\omega^\mathrm{2PN}\leq -0.00045^\circ\,\mathrm{\,yr}^{-1}$. A recently released prediction of $\dot\omega^\mathrm{2PN}$ for \psrab\, is discussed.
\end{abstract}

{
\textit{Unified Astronomy Thesaurus concepts}:   Gravitation\,(661); General relativity\,(641);  Relativistic mechanics\,(1391); Neutron stars(1108)
}


\section{Introduction}
Recently, \citet{2020MNRAS.497.3118H} performed a detailed analysis of the perspectives of measuring, or effectively constraining, the moment of inertia (MOI) $\mathcal{I}_\mathrm{A}$ of the pulsar \psra\, \citep{2003Natur.426..531B,2004Sci...303.1153L} by the end of the present decade  by exploiting the general relativistic spin--orbit Lense--Thirring periastron precession $\dot\omega^\mathrm{LT,\,A}$ \citep{1988NCimB.101..127D} induced by its spin angular momentum ${\bds S}^\mathrm{A}$. Among the competing dynamical effects acting as potential sources of systematic uncertainty, \citet{2020MNRAS.497.3118H} included also the periastron precession $\dot\omega^\mathrm{2PN}$ to the second post-Newtonian (2PN) order which, along with the much larger\footnote{Also $\dot\omega^\mathrm{LT}$ is a 1PN effect because it is proportional to $c^{-2}$.} 1PN
precession
\eqi
\dot\omega^\mathrm{1PN} = \rp{3\,n^\mathrm{K}\,\mu}{c^2\,a\,\ton{1-e^2}},\lb{om1}
\eqf
depends only on the masses $M_\mathrm{A},\,M_\mathrm{B}$ of both the neutron stars which the Double Pulsar \psrab\, is made of.
In \rfr{om1}, $c$ is the speed of light in vacuum, $\mu\doteq GM$ is the gravitational parameter of the Double Pulsar  made of the product of the Newtonian constant of gravitation $G$ times the sum of the masses $M\doteq M_\mathrm{A}+M_\mathrm{B}$, $a$ and $e$ are  the osculating
numerical values of the semimajor axis and eccentricity, respectively, at the same arbitrary moment of time $t_0$ \citep{1994ApJ...427..951K}, while
\eqi
n^\mathrm{K}\doteq\sqrt{\rp{\mu}{a^3}}\lb{nK}
\eqf
is the osculating Keplerian mean motion.
In particular, \citet[][Table\,1]{2020MNRAS.497.3118H} reported
\eqi
\dot\omega^\mathrm{2PN} = 0.000439^\circ\,\mathrm{yr}^{-1} = 1.58\,\mathrm{''\,yr}^{-1},\lb{vaffa}
\eqf
for the 2PN periastron precession which would, thus, be prograde. In \rfr{vaffa}, $\mathrm{''}$ stands for arcseconds. \Rfr{vaffa} is to be compared with the retrograde Lense--Thirring periastron rate which,
if calculated with the latest determination of $\mathcal{I}_\mathrm{A}$ by \citet{2021PhRvL.126r1101S},
would be of comparable magnitude
\eqi
\dot\omega^\mathrm{LT,\,A} \simeq  -0.0006^\circ\,\mathrm{yr}^{-1} = -2.16\,\mathrm{''\,yr}^{-1}.\lb{oLT}
\eqf
It is clear that an accurate prediction of the 2PN periastron precession, or of the experimental quantity related to it which is actually determined in real data analyses, is of the utmost importance since, according to \rfr{vaffa}, it may cancel \rfr{oLT} to a large extent. To this aim, it is important to stress that, although seemingly unnoticed so far in the literature, a certain amount of uncertainty should be deemed as still lingering on that matter because, perhaps, of how $\dot\omega$ is routinely expressed in most of the papers devoted to binary pulsars. Indeed, as it will be shown here, the way usually adopted in the literature to write the total 2PN periastron precession hides its dependence on the initial conditions which, indeed, is buried in the parameterization used. Such a distinctive feature does not occur at the 1PN level whose averaged orbital precessions such as \rfr{om1} and the Lense--Thirring one $\dot\omega^\mathrm{LT}$ are independent of the orbital phase at a reference epoch. Thus, while the predictions of the 1PN precessions are valid for any starting time, it is not so for the 2PN ones, despite their--purely formal--independence of the initial conditions in certain parameterizations. Moreover, there is some confusion about the periastron \textit{precession} and how to correctly calculate it from the fractional periastron \textit{advance per orbit}. Finally, in numerically calculating $\dot\omega^\mathrm{2PN}$, the fact that also the formal 1PN term contributes it in a subtle way is often overlooked yielding wrong results.

The paper is organized as follows.  In Section\,\ref{sec.2}, the total 2PN periastron rate is calculated (see \rfr{super}) by using the osculating Keplerian orbital elements from existing expressions in the literature for the fractional PN periastron shift per orbit $k^\mathrm{1PN}+k^\mathrm{2PN}$. In particular, Equation\,(21) by \citet{2021Univ....7...37I} is used as starting point in Section\,\ref{sec.2.1}, where an error by \citet{2021Univ....7...37I} in obtaining the true total 2PN periastron rate is disclosed and corrected. In Section\,\ref{sec.2.2}, \rfr{super} is obtained starting from Equation\,(5.18) by \citet{1988NCimB.101..127D}, expressed in the DD parameterization, after a proper conversion from the latter to the osculating Keplerian orbital elements. In Section\,\ref{sec.3}, \rfr{super} is confirmed by numerically integrating the PN equations of motion up to the 2PN order  for a fictitious binary system. The results of Section\,\ref{sec.2} are applied to other astrophysical and astronomical systems of interest in Section\,\ref{sec.4}. The case of \psrab\, is dealt with in Section\,\ref{sec.4.1}, where \rfr{vaffa} is discussed as well. Section\,\ref{sec.4.2} treats Mercury, the spacecraft Juno orbiting Jupiter, the Earth's artificial satellites LAGEOS II, and the S-star S4711 around Sgr A$^\ast$. Section\,\ref{sec.5} summarizes the findings obtained and offers concluding remarks.
\section{How to correctly calculate the 2PN periastron precession from the fractional 1PN+2PN periastron shift per orbit using the osculating Keplerian orbital elements}\lb{sec.2}
In pulsar timing analyses, one of the so called post-Keplerian (PK) parameters which are determined for several binary pulsars is the fractional periastron shift per orbit $k^\mathrm{PK}$ defined as
\eqi
k^\mathrm{PK}\doteq \rp{\ang{\Delta\omega^\mathrm{PK}}}{2\uppi}. \lb{kap}
\eqf
In \rfr{kap}, $\omega$ is the argument of periastron, $\Delta\omega^\mathrm{PK}$ is the time-dependent shift of periastron induced by some PK dynamical extra--acceleration  with respect to the Newtonian inverse-square one, and the angular brackets $\ang{\cdots}$ denote the average over the orbital period $P^\mathrm{PK}$ which, in presence of PK accelerations, has to be meant as the anomalistic period $P_\mathrm{ano}^\mathrm{PK}$, i.e. the time span between two successive crossings of the (moving) periastron position.
Nonetheless, it is common practice to deal with the averaged\footnote{In the following, the brackets $\ang{\cdots}$ around $\dot\omega^\mathrm{PK}$ will be omitted in order to make the notation less cumbersome.}  periastron precession $\dot\omega^\mathrm{PK}$ which is connected with $k^\mathrm{PK}$ through
\eqi
k^\mathrm{PK} = \rp{\dot\omega^\mathrm{PK}}{n^\mathrm{PK}},
\eqf
where
\eqi
n^\mathrm{PK}\doteq \rp{2\uppi}{P^\mathrm{PK}}\lb{nPK}
\eqf
is the PK mean motion. In general, \rfr{nPK} differs from \rfr{nK}.
\subsection{Starting from the formula by Iorio in osculating Keplerian orbital elements}\lb{sec.2.1}
As far as the 2PN periastron advance is concerned, \citet[][Equation\,(21)]{2021Univ....7...37I}
correctly calculated $k^\mathrm{2PN}$, up to the scaling factor $n^\mathrm{K}$, with the Gauss perturbing equations in terms of the osculating Keplerian orbital elements (see \rfr{o2pn}) by showing that his expression agrees with those obtained by \citet{1994ARep...38..104K} with the same perturbative technique but a different calculational strategy, and by \citet{1988NCimB.101..127D} who, instead, used the Hamilton-Jacobi method and the Damour-Deruelle (DD) parameterization \citep{1985AIHS...43..107D} which is nowadays routinely used in standard pulsar timing analyses \citet{1986AIHS...44..263D,1992PhRvD..45.1840D}.  \citet{2021Univ....7...37I}, after having scaled $k^\mathrm{2PN}$ by \rfr{nK}, erroneously claimed that the resulting expression for $n^\mathrm{K}\,k^\mathrm{2PN}$ is the \textit{total} 2PN pericentre \textit{precession}, which is not the case, as it will be shown below.
Here, the explicit expressions of $k^\mathrm{1PN},\,k^\mathrm{2PN}$ in terms of the osculating Keplerian orbital elements are reported. They are
\begin{align}
k^\mathrm{1PN}&=\rp{3\,\mu}{c^2\,a\,\ton{1-e^2}}\lb{o1pn}, \\ \nonumber \\
k^\mathrm{2PN}&=\rp{3\,\mu^2\,\qua{2-4\,\nu +e^2\,\ton{1+10\,\nu} + 16\,e\,\ton{-2+\nu}\,\cos f_0 }}{4\,c^4\,a^2\,\ton{1-e^2}^2},\lb{o2pn}
\end{align}
where $f_0$ is the osculating numerical value of the true anomaly $f$ at  some arbitrary moment of time $t_0$, and
\eqi
\nu\doteq \rp{M_\mathrm{A}\,M_\mathrm{B}}{M^2}.
\eqf

In order to correctly calculate the \textit{total} 2PN  pericentre \textit{precession} $\dot\omega^\mathrm{2PN}$, some characteristic time interval playing the role of \virg{orbital period} has to be worked out to the 1PN order.
In the present case, the anomalistic period, i.e. the time interval between two successive crossings of the (moving) pericentre position, fulfils such a requirement.
To the 1PN order, it can be written as
\eqi
P^\mathrm{1PN}_\mathrm{ano} = P^\mathrm{K} + \Delta P_\mathrm{ano}^\mathrm{1PN},
\eqf
where the osculating Keplerian period is
\eqi
P^\mathrm{K}\doteq \rp{2\uppi}{n^\mathrm{K}}=2\uppi\sqrt{\rp{a^3}{\mu}},
\eqf
and the 1PN correction, calculated according to the strategy followed by \citet{2016MNRAS.460.2445I}, turns out to be
\eqi
\Delta P_\mathrm{ano}^\mathrm{1PN} = \rp{\uppi\,\sqrt{a\,\mu}}{2\,c^2\,\ton{1-e^2}^2}\,\mathcal{T}_\mathrm{ano}^\mathrm{1PN},
\eqf
with
\begin{align}
\mathcal{T}_\mathrm{ano}^\mathrm{1PN}\nonumber & \doteq 36 + e^2\,\ton{42 - 38\,\nu} + 2\,e^4\,\ton{6 - 7\,\nu} - 8\,\nu + \\ \nonumber \\
& + 3\,e\,\grf{\qua{28 + 3\,e^2\,\ton{4 - 5\,\nu} - 12\,\nu}\,\cos f_0 - e\,\ton{-10 + 8\,\nu + e\,\nu\,\cos f_0}\,\cos 2f_0}. \lb{tpn}
\end{align}
In the point particle limit corresponding to $\nu\rightarrow 0$, \rfr{tpn} reduces to Equation\,(72) of \citet{2016MNRAS.460.2445I}.
The 1PN mean motion is, thus,
\eqi
n^\mathrm{1PN}\doteq\rp{2\uppi}{P_\mathrm{ano}^\mathrm{1PN}}=\rp{n^\mathrm{K}}{1+\rp{\mu}{4\,c^2\,a\,\ton{1-e^2}^2}\,\mathcal{T}_\mathrm{ano}^\mathrm{1PN} }.\lb{nPN}
\eqf

Note that Equation\,(21) of \citet{2021Univ....7...37I}, i.e. the product of \rfr{nK} times \rfr{o2pn}
\eqi
n^\mathrm{K}\,k^\mathrm{2PN}=\rp{3\,\mu^{5/2}\,\qua{2-4\,\nu +e^2\,\ton{1+10\,\nu} + 16\,e\,\ton{-2+\nu}\,\cos f_0 }}{4\,c^4\,a^{7/2}\,\ton{1-e^2}^2},\lb{cazz}
\eqf
has formally the dimensions of a pericentre \textit{precession} of the order of $\mathcal{O}\ton{c^{-4}}$, but, contrary to what mistakenly claimed by \citet{2021Univ....7...37I}, it is \textit{not} the \textit{total} 2PN pericentre \textit{rate} $\dot\omega^\mathrm{2PN}$.
Indeed, the correct analytical expression for it can only be obtained by retaining the term of the order of $\mathcal{O}\ton{c^{-4}}$ in the  expansion in powers of $c^{-1}$ of the product of \rfr{nPN} times the sum of \rfr{o1pn} and \rfr{o2pn}. If, on the one hand, replacing $n^\mathrm{K}$ with \rfr{nPN} does not affect \rfr{cazz} in the power expansion to the 2PN order, on the other hand, it does matter when it is \rfr{o1pn} that is multiplied by \rfr{nPN} and power-expanded to the order of $\mathcal{O}\ton{c^{-4}}$. Indeed, from \rfr{o1pn} and \rfr{nPN} one has
\begin{align}
\left. n^\mathrm{1PN}\,k^\mathrm{1PN}\right|_\mathrm{2PN} \nonumber &= \rp{3\,\mu^{5/2}}{4\,c^4\,a^{7/2}\,\ton{1-e^2}^3}\,\ton{
-36 + 8\,\nu + 2\,e^4\,\ton{-6 + 7\,\nu} + e^2\,\ton{-42 + 38\,\nu} + \right.\\ \nonumber \\
\nonumber &+\left. 3\,e\,\grf{\qua{4\,\ton{-7 + 3\,\nu} + 3\,e^2\,\ton{-4 + 5\,\nu}}\,\cos f_0 + \right.\right.\\ \nonumber \\
&+\left.\left. e\,\ton{-10 + 8\,\nu + e\,\nu\,\cos f_0}\,\cos 2 f_0}
},
\end{align}
which, added to \rfr{cazz}, yields
\begin{align}
\dot \omega^\mathrm{2PN} \nonumber & =\rp{3\,\mu^{5/2}}{8\,c^4\,a^{7/2}\,\ton{1-e^2}^3}\,\grf{
-68 + 8\,\nu + e^4\,\ton{-26 + 8\,\nu} + 2\,e^2\,\ton{-43 + 52\,\nu} + \right.\\ \nonumber \\
\nonumber &+\left. e\,\qua{8\,\ton{-29 + 13\,\nu} + e^2\,\ton{-8 + 61\,\nu}}\,\cos f_0 + \right.\\ \nonumber\\
&+\left. 3\,e^2\,\qua{4\,\ton{-5 + 4\,\nu}\,\cos 2f_0 + e\,\nu\,\cos 3f_0}\lb{super}
}.
\end{align}
This is the right analytical expression for the \textit{full} 2PN pericentre \textit{precession} expressed in terms of the osculating Keplerian orbital elements.

Recapitulating, on the one hand, \citet[][Equation\,(21)]{2021Univ....7...37I} correctly worked out the 2PN fractional pericentre shift per orbit $k^\mathrm{2PN}$ up to $n^\mathrm{K}$ as  scaling factor.  On the other hand,  \citet{2021Univ....7...37I}, after having multiplied it by the osculating Keplerian mean motion $n^\mathrm{K}$, mistakenly claimed that the resulting expression for $n^\mathrm{K}\,k^\mathrm{2PN}$ was the \textit{total} 2PN pericentre \textit{precession}, missing a further contribution from the power expansion to the 2PN order of the product $n^\mathrm{1PN}\,k^\mathrm{1PN}$.
\subsection{Starting from the formula by Damour and Sch{\"a}fer in the Damour-Deruelle parameterization}\lb{sec.2.2}
\Rfr{super} is in agreement also with the expression for the total 2PN pericentre precession, written in terms of the osculating Keplerian orbital elements, which can be extracted from
\citep[][Equation\,(5.18)]{1988NCimB.101..127D}
\begin{align}
k^\mathrm{1PN} + k^\mathrm{2PN} \nonumber \lb{kDD}&= \rp{3\,\ton{\mu\,n_\mathrm{DD}}^{2/3}}{c^2\,\ton{1-e_\mathrm{T}^2}}\,\qua{1+ \rp{\ton{\mu\,n_\mathrm{DD}}^{2/3}}{c^2\,\ton{1-e_\mathrm{T}^2}}\ton{\rp{39}{4}x_\mathrm{A}^2 + \rp{27}{4}x^2_\mathrm{B} + 15\,x_\mathrm{A}\,x_\mathrm{B}} - \right.\\ \nonumber \\
&\left. - \rp{\ton{\mu\,n_\mathrm{DD}}^{2/3}}{c^2}\,\ton{\rp{13}{4}x_\mathrm{A}^2 + \rp{1}{4}x^2_\mathrm{B} + \rp{13}{3}\,x_\mathrm{A}\,x_\mathrm{B}} }.
\end{align}
In \rfr{kDD},
\begin{align}
x_\mathrm{A}&\doteq \rp{M_\mathrm{A}}{M}, \\ \nonumber\\
x_\mathrm{B}&\doteq \rp{M_\mathrm{B}}{M}=1-x_\mathrm{A},
\end{align}
while $e_\mathrm{T}$ and $n_\mathrm{DD}$ are members of the Damour-Deruelle (DD) formalism \citep{1985AIHS...43..107D} which, in the limit $c\rightarrow\infty$, reduce to the Keplerian eccentricity $e$ and mean motion $n^\mathrm{K}$, as it will be shown below.

The \virg{proper time} eccentricity $e_\mathrm{T}$ reads \citep[][pag.\,272]{1986AIHS...44..263D}
\eqi
e_\mathrm{T} = e_t\,\ton{1+\delta} + e_\theta - e_r,\lb{eTi}
\eqf
where \citep[][Equation\,(3.8b)]{1985AIHS...43..107D}
\eqi
e_t=\rp{e_R}{1+\rp{\mu}{c^2\,a_R}\,\ton{4-\rp{3}{2}\,\nu}},\lb{noo}
\eqf
\citep[][Equation\,(4.13)]{1985AIHS...43..107D}
\eqi
e_\theta = e_R\,\ton{1+\rp{\mu}{2\,c^2\,a_R}},
\eqf
\citep[][Equation\,(6.3b)]{1985AIHS...43..107D}
\eqi
e_r=e_R\,\qua{1-\rp{\mu}{2\,c^2\,a_R}\,\ton{x^2_\mathrm{A}-\nu}},
\eqf
and
\citep[][Equation\,(20)]{1986AIHS...44..263D}
\eqi
\delta=\rp{\mu}{c^2\,a_R}\,\ton{x_\mathrm{A}\,x_\mathrm{B} + 2\,x_\mathrm{B}^2}.\lb{sii}
\eqf
In \rfrs{noo}{sii}, $a_R$ is another member of the DD parameterization.
According to \rfrs{noo}{sii}, \rfr{eTi} can be expressed in terms of only $a_R,\,e_R$ as
\eqi
\rp{e_\mathrm{T}}{e_R} = \rp{1 + \rp{\mu}{2\,c^2\,a_R}\,\qua{4+3\,\ton{x_\mathrm{A}-2}\,x_\mathrm{A}} + \rp{\mu^2}{4\,c^4\,a^2_R}\,\ton{8-3\,\nu}\,x^2_\mathrm{A} }{1+\rp{\mu}{2\,c^2\,a_R}\,\ton{8-3\,\nu}}.\lb{eTTi}
\eqf
The DD mean motion is \citep[][Equation\,(3.7)]{1985AIHS...43..107D}
\eqi
n_\mathrm{DD} \doteq \sqrt{\rp{\mu}{a^3_R}}\,\qua{1+ \rp{\mu}{2\,c^2\,a_R}\,\ton{-9+\nu}}.\lb{nDD}
\eqf
\Rfrs{eTTi}{nDD} are both functions of $a_R,\,e_R$ which, in turn, can be expressed in terms of the osculating Keplerian semimajor axis $a$ and eccentricity $e$ by means of
\citep[][Equations\,(28)-(29)]{1994ApJ...427..951K}
\begin{align}
a_R \lb{conva}& = a- da_0 - \rp{\mu}{c^2\,\ton{1-e^2}^2}\,\qua{
-3 + \nu + e^2\,\ton{-13 + e^2 + 7\,\nu + 2\,e^2\,\nu}
}, \\ \nonumber \\
e_R \lb{conve} & = e -de_0 -\rp{e\,\mu}{2\,c^2\,a\,\ton{1-e^2}}\,\qua{-17 + 6\,\nu + e^2\,\ton{2 + 4\,\nu}},
\end{align}
with
\citep[][Equation\,(14)]{1994ApJ...427..951K}
\begin{align}
da_0 \lb{da0} \nonumber &= \rp{e\,\mu}{4\,c^2\,\ton{1-e^2}^2}\,\grf{
\qua{8\,\ton{-7 + 3\,\nu} + e^2\,\ton{-24 + 31\,\nu}}\,\cos f_0 + \right.\\ \nonumber \\
&\left. + e\,\qua{4 \ton{-5 + 4\,\nu}\,\cos 2f_0 + e\,\nu\,\cos 3f_0}
},
\end{align}
and \citep[][Equation\,(16)]{1994ApJ...427..951K}
\begin{align}
de_0 \lb{de0} \nonumber &= \rp{\mu}{8\,c^2\,a\,\ton{1-e^2}}\,\grf{
\qua{8\,\ton{-3 + \nu} + e^2\,\ton{-56 + 47 \nu}}\,\cos f_0  + \right.\\ \nonumber \\
&\left. + e\,\qua{4 \ton{-5 + 4 \nu}\,\cos 2 f_0  + e\,\nu\,\cos 3 f_0 }
}.
\end{align}
Note that \rfrs{conva}{de0} are written for general relativity; their general expressions for a given class of alternative theories of gravitation can be found in \citet{1994ApJ...427..951K}.
The final expressions  for $a_R,\,e_R$ are
\begin{align}
\rp{a_R}{a} \nonumber \lb{aR}& = 1 - \rp{\mu}{c^2\,a\,\ton{1-e^2}^2}\,\qua{-3 + \nu + e^4\,\ton{1 + 2\,\nu} + e^2\,\ton{-13 + 7\,\nu}} + \\ \nonumber \\
\nonumber & + e\,\rp{\mu}{4\,c^2\,a\,\ton{1-e^2}^2}\,\grf{\qua{56 + e^2\,\ton{24 - 31\,\nu} - 24\,\nu}\,\cos f_0 + \right.\\ \nonumber \\
&\left. + e\,\qua{4\,\ton{5 - 4\,\nu}\,\cos 2 f_0 - e\,\nu\,\cos 3 f_0}}, \\ \nonumber \\
\rp{e_R}{e} \nonumber \lb{eR}& = 1 - \rp{\mu}{2\,c^2\,a\ton{1-e^2}}\,\qua{-17 + 6\,\nu + e^2\,\ton{2 + 4\,\nu}} + \\ \nonumber \\
\nonumber & - \rp{\mu}{8\,c^2\,a\,e\,\ton{1-e^2}}\,\grf{\qua{8\,\ton{-3 + \nu} + e^2\,\ton{-56 + 47\,\nu}}\,\cos f_0  + \right. \\ \nonumber \\
&+\left. e\,\qua{4\,\ton{-5 + 4\,\nu}\,\cos 2 f_0  + e\,\nu\,\cos 3 f_0}}.
\end{align}
By using  \rfrs{aR}{eR}, \rfrs{eTTi}{nDD}  can  be finally expressed, to the order of $\mathcal{O}\ton{c^{-2}}$, as
\begin{align}
%
%
%
%
%
\rp{8\,c^2\,a\,\ton{e-e_\mathrm{T}}\,\ton{1-e^2}}{\mu} \nonumber \lb{due}& = \qua{8\,\ton{-3 + \nu} + e^2\,\ton{-56 + 47\,\nu}}\cos f_0  + \\ \nonumber \\
\nonumber & + e\,\ton{4\,\grf{-13 + 3\,\nu - 3\,\ton{-2 + x_\mathrm{A}} x_\mathrm{A} + \right.\right.\\ \nonumber \\
\nonumber &\left.\left. + e^2\,\qua{-2 + 7\,\nu + 3\,\ton{-2 + x_\mathrm{A}} x_\mathrm{A}}} + \right. \\ \nonumber \\
&\left. + 4\,\ton{-5 + 4\,\nu}\cos 2f_0  + e\,\nu\cos 3f_0}, \\ \nonumber \\
\ton{\rp{n_\mathrm{DD}}{n^\mathrm{K}}-1}\,\rp{8\,c^2\,a\,\ton{1-e^2}^2}{\mu} \nonumber \lb{tre}& =
8\,\ton{-9 + 2\,\nu} + 4\,e^4\,\ton{-6 + 7\,\nu} + e^2\,\ton{-84 + 76\,\nu} + \\ \nonumber \\
\nonumber & + 3\,e\,\grf{\qua{8\,\ton{-7 + 3\,\nu} + e^2\,\ton{-24 + 31\,\nu}}\cos f_0  + \right. \\ \nonumber \\
&\left. + e\,\qua{4\,\ton{-5 + 4\,\nu}\cos 2f_0  + e\,\nu\cos 3f_0}}.
\end{align}
A power expansion to the order of $\mathcal{O}\ton{c^{-4}}$ of the product of  \rfr{nDD} by \rfr{kDD}, calculated with \rfrs{due}{tre}, yields just \rfr{super}.

About the fractional periastron shift per orbit,  \citet{2021Univ....7...37I} demonstrated the equivalence of his Equation\,(21), up to the scaling factor $n^\mathrm{K}$, with the \textit{total} \textit{explicit} components of the order of $\mathcal{O}\ton{c^{-4}}$ of \rfr{kDD} and of an analogous formula by \citet{1994ARep...38..104K}, once the proper translation of the latter ones into the osculating Keplerian orbital elements was appropriately carried out.
\section{Numerically integrating the 1PN+2PN equations of motion}
\begin{figure}[H]
\centering
\centerline{
\vbox{
\begin{tabular}{c}
\epsfxsize= 14 cm\epsfbox{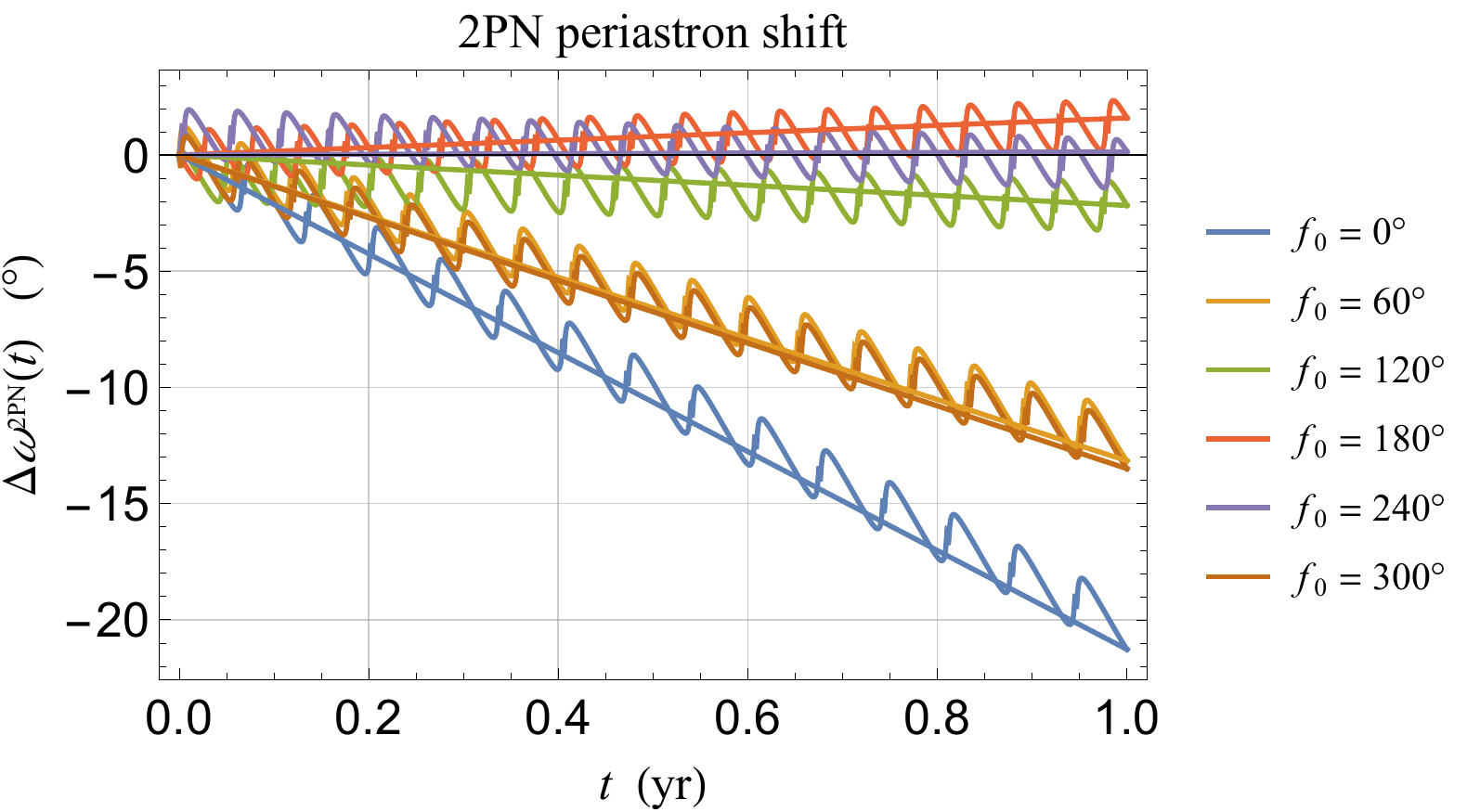}\\
\epsfxsize= 14 cm\epsfbox{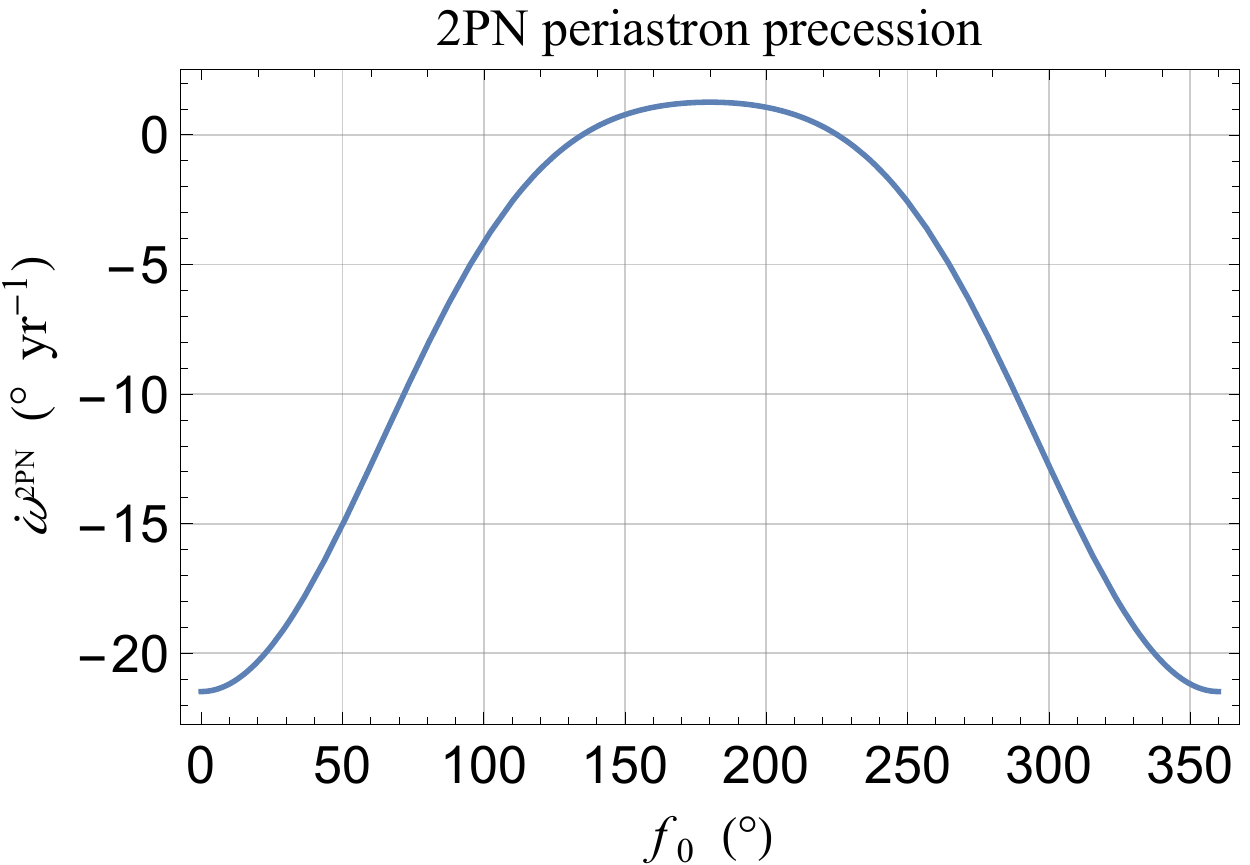}\\
\end{tabular}
}
}
\caption{
Upper panel: Time-dependent signatures $\Delta\omega^\mathrm{2PN}\ton{t}$ (see the text for details on their generation) obtained by numerically integrating the 1PN+2PN equations of motion for a fictitious two-body system with $M_\mathrm{A}= 1\times 10^6\,M_\odot,\,M_\mathrm{B} = 2\times 10^6\,M_\odot,\,\Pb=0.05\,\mathrm{yr},\,e=0.75$ for different values of the true anomaly at epoch $f_0$. The superimposed straight lines are linear fits to the corresponding $\Delta\omega^\mathrm{2PN}\ton{t}$. The units are $^\circ$. Lower panel: Plot of the 2PN periastron precession, in $^\circ\,\mathrm{yr}^{-1}$, analytically calculated for the same binary system with \rfr{super} as a function of $f_0$.
}\label{fig1}
\end{figure}

\lb{sec.3}
The correctness of \rfr{super}, and also its general applicability to whatsoever binary system for which the PN approximation is deemed applicable, can be numerically demonstrated in the following way.
For the sake of clarity, a fictitious two-body system made of, say,  two supermassive black holes with $M_\mathrm{A}= 1\times 10^6\,M_\odot,\,M_\mathrm{B} = 2\times 10^6\,M_\odot$ orbiting along a highly eccentric ($e=0.75$) orbit in $0.05\,\mathrm{yr}$ is considered.
For a given set of initial conditions, parameterized in terms of the\footnote{Actually, such a choice is, by no means, necessary, being any other one yielding a bound trajectory equally valid; in any case, by suitably varying the initial conditions, the resulting time series for the periastron evolution would change their slopes.} Keplerian orbital elements, the equations of motion, in rectangular Cartesian harmonic coordinates, including the PN accelerations
\citetext{\citealp[see, e.g.,][Eq.\,(4.4.28),\,p.\,154]{1991ercm.book.....B}; \citealp[Eq.\,(A2.6),\,p.\,166]{Sof89}; \citealp[Eq.\,(10.3.7),\,p.\,381]{SoffelHan19}}
\begin{align}
{\bds A}^\mathrm{1PN} &= \rp{\mu}{c^2\,r^2}\grf{\qua{\ton{4+2\,\nu}\,\rp{\mu}{r} + \rp{3}{2}\,\nu\,{\mathrm{v}}_r^2-\ton{1+3\,\nu}\,\mathrm{v}^2}\,\bds{\hat{r}}+ \ton{4-2\,\nu}\,{\mathrm{v}}_r\,{\mathbf{v}}}\lb{mega1PN},
\end{align}
and
\citetext{\citealp[see, e.g.,][Eq.\,(4.4.29),\,p.\,154]{1991ercm.book.....B}; \citealp[Eq.\,(2.2d),\,p.\,825]{1995PhRvD..52..821K};  \citealp[Eq.\,(B11),\,p.\,10]{PhysRevD.82.104031}}
\begin{align}
{\bds A}^\mathrm{2PN} \nonumber & = \rp{\mu}{c^4\,r^2}\grf{
\qua{\nu\,\ton{-3 + 4\,\nu}\,\mathrm{v}^4 + \rp{15}{8}\,\nu\,\ton{-1 + 3\,\nu}\,{\mathrm{v}}_r^4 +\,\nu\,\ton{\rp{9}{2} - 6\,\nu}\,\mathrm{v}^2\,{\mathrm{v}}_r^2 +\,\nu\,\ton{\rp{13}{2} - 2\,\nu}\,\rp{\mu}{r}\,\mathrm{v}^2 + \right.\right.\\ \nonumber \\
\nonumber &\left.\left. + \ton{2 + 25\,\nu + 2\,\nu^2}\,\rp{\mu}{r}\,{\mathrm{v}}_r^2 -\ton{9 + \rp{87}{4}\,\nu}\,\rp{\mu^2}{r^2}
}\,\bds{\hat{r}}
+\qua{
\nu\,\ton{\rp{15}{2} + 2\,\nu}\,\mathrm{v}^2 -\nu\,\ton{\rp{9}{2} + 3\,\nu}\,{\mathrm{v}}_r^2 - \right.\right.\\ \nonumber \\
&\left.\left. - \ton{2 + \rp{41}{2}\,\nu + 4\,\nu^2}\,\rp{\mu}{r}
}\,{\mathrm{v}}_r\,{\mathbf{v}}}\lb{mega2PN}
\end{align}
in addition to the Newtonian monopole
\eqi
{\bds{A}}^\mathrm{N} = -\rp{\mu}{r^2}\,\bds{\hat{r}},\lb{megaN}
\eqf
are numerically integrated over 1 yr with and without \rfrs{mega1PN}{mega2PN} in each run, and time series of $\omega\ton{t}$ are correspondingly calculated. In \rfrs{mega1PN}{megaN}, $r$ is the relative distance between A and B, $\bds{\hat{r}}$ is the versor of the relative position vector,  $\boldsymbol{\mathrm{v}}$ is the velocity vector of the relative motion, and $\mathrm{v}_r\doteq\boldsymbol{\mathrm{v}}\boldsymbol\cdot\bds{\hat{r}}$ is the radial velocity.   Then, the difference between the Newtonian and the PN time series for $\omega\ton{t}$ is taken to extract the time-dependent PN shift $\Delta\omega^\mathrm{PN}\ton{t}$. By construction, it includes both the full 1PN and 2PN contributions along with other terms of higher order due to the interplay between the 1PN and 2PN accelerations, not of interest here. In order to single out just the total 2PN effect (up to other PN contributions of higher order) $\Delta\omega^\mathrm{2PN}\ton{t}$, the 1PN linear trend, analytically calculated by multiplying $k^\mathrm{1PN}$ of \rfr{o1pn} times $n^\mathrm{K}\,t$, is subtracted from the time series $\Delta\omega^\mathrm{PN}\ton{t}$. The same procedure is repeated by varying the true anomaly at epoch $f_0$ leading to a change of the initial conditions. The resulting time-dependent signatures for $\Delta\omega^\mathrm{2PN}\ton{t}$ are displayed in the upper panel of Figure\,\ref{fig1}. A linear fit to each of them is performed, and the resulting straight lines are superimposed.  Their slopes, in $^\circ\,\mathrm{yr}^{-1}$, can be compared with the lower panel of Figure\,\ref{fig1} displaying the plot of \rfr{super} as a function of $f_0$; the agreement is neat.
\section{Application to the Double Pulsar and other systems}\lb{sec.4}
\subsection{The case of \psrab}
\begin{figure}[H]
\centering
\centerline{
\vbox{
\begin{tabular}{c}
\epsfxsize= 16 cm\epsfbox{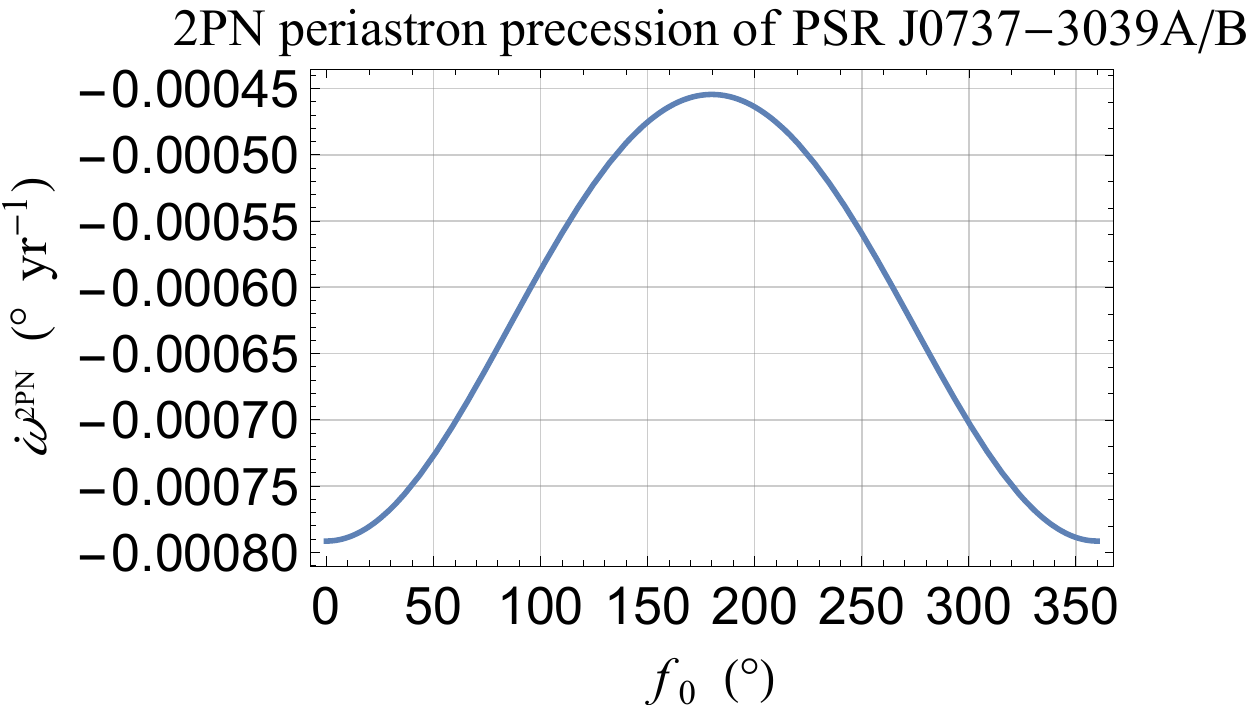}\\
\end{tabular}
}
}
\caption{
Plot of the total 2PN periastron precession of \psrab, in $^\circ\,\mathrm{yr}^{-1}$, analytically calculated with \rfr{super} as a function of $f_0$.
}\label{fig2}
\end{figure}

\lb{sec.4.1}
In the case of \psrab, \rfr{super} yields
\eqi
-0.00080^\circ\,\mathrm{yr}^{-1} \leq\dot\omega^\mathrm{2PN}\leq -0.00045^\circ\,\mathrm{\,yr}^{-1},\lb{merda1}
\eqf
as shown by Figure\,\ref{fig2},
or, equivalently,
\eqi
-2.8\,\mathrm{''\,yr}^{-1}\leq\dot\omega^\mathrm{2PN}\leq  -1.6\,\mathrm{''\,yr}^{-1}.\lb{merda2}
\eqf
\Rfrs{merda1}{merda2} correct the wrong range  for $\dot\omega^\mathrm{2PN}$ which one would obtain for the Double Pulsar by summing the values in Equation\,(20) and Equation\,(21) of\footnote{Also the remaining numerical results in  \citet[][Section\,2,\,pag.\,6]{2021Univ....7...37I} are wrong, and should be calculated with \rfr{super} for each of the other binary systems considered.} \citet{2021Univ....7...37I}. From Figure\,\ref{fig2}, it can be noted that the 2PN periastron precession of \psrab\, is always retrograde and does not vanish for any value of $f_0$.

About \rfr{vaffa} quoted by \citet[][Table\,1]{2020MNRAS.497.3118H}, it was obtained as follows.
By taking the product of \rfr{nDD}, or, in this case, also of \rfr{nK}, times the sum of only the second and the third term in\footnote{Equation\,(2) of \citet{2020MNRAS.497.3118H}, up to the spin--orbit term, is just the product of \rfr{kDD}, written in terms of some \virg{orbital frequency} $\nk$, times $\nk$ itself. In particular, $f_\mathrm{O}$ entering Equation\,(2) of \citet{2020MNRAS.497.3118H} is not to be confused with the true anomaly at epoch $f_0$.} \rfr{kDD} and expanding the resulting expression to the order of $\mathcal{O}\ton{c^{-4}}$, one gets the following quantity which is dimensionally a 2PN precession
\eqi
\dot\psi^\mathrm{2PN}=\rp{n^\mathrm{K}\,\mu^2\,\grf{78-28\,\nu + e^2\,\qua{3 + 2\,\ton{23-5\,x_\mathrm{A}}\,x_\mathrm{A}} }}{4\,c^4\,a^2\,\ton{1-e^2}^2}=0.000439^\circ\,\mathrm{yr}^{-1},
\eqf
in agreement with \rfr{vaffa}. This demonstrates that the prediction for the 2PN periastron precession  by \citet[][Table\,1]{2020MNRAS.497.3118H}
is, in fact, incomplete since it neglected the contribution to $\dot\omega^\mathrm{2PN}$ of the product of \rfr{nDD} times the first term in \rfr{kDD}, which is formally of the order of $\mathcal{O}\ton{c^{-2}}$.
\subsection{Other astronomical systems}\lb{sec.4.2}
For the Sun and Mercury, \rfr{super} yields
\eqi
-18\,\upmu\mathrm{as\,cty}^{-1}\leq\dot\omega_{\mercury}^\mathrm{2PN}\leq  -4\,\upmu\mathrm{as\,cty}^{-1},\lb{merc}
\eqf
while for the spacecraft Juno currently orbiting Jupiter, it is
\eqi
-4\,\upmu\mathrm{as\,yr}^{-1}\leq\dot\omega^\mathrm{2PN}\leq  0\,\upmu\mathrm{as\,yr}^{-1}.\lb{jun}
\eqf
The 2PN perigee precession of the Earth's artificial satellite LAGEOS II is as little as
\eqi
-0.0108\,\upmu\mathrm{as\,yr}^{-1}\leq\dot\omega^\mathrm{2PN}\leq  -0.0100\,\upmu\mathrm{as\,yr}^{-1}.\lb{lag}
\eqf
In \rfrs{merc}{lag}, $\upmu\mathrm{as}$ stands for microarcseconds.

Larger values occur, e.g., for the recently discovered S-star S4711 \citep{2020ApJ...899...50P} orbiting the supermassive black hole  in the Galactic Center at Sgr A$^\ast$; it revolves
around its primary in $7.6\,\mathrm{yr}$ along an orbit with an eccentricity as large as $e=0.768$. Its 2PN periastron precession range turns out to be
\eqi
-1.4\,\mathrm{''\,yr}^{-1}\leq\dot\omega^\mathrm{2PN}\leq  0.074\,\mathrm{''\,yr}^{-1}.\lb{S4711}
\eqf
%
%
\section{Summary and conclusions}\lb{sec.5}
Calculating correctly the \textit{total} 2PN periastron \textit{precession} $\dot\omega^\mathrm{2PN}$ from the fractional periastron \textit{advance per orbit} $k^\mathrm{PN}=k^\mathrm{1PN}+k^\mathrm{2PN}$ requires to multiply the latter one by the 1PN mean motion $n^\mathrm{1PN}$ instead of the osculating Keplerian one $n^\mathrm{K}$, as incorrectly done by \citet{2021Univ....7...37I}, and to expand  the resulting expression to the order of $\mathcal{O}\ton{c^{-4}}$. It remains true independently of the parameterization used. Adopting the osculating Keplerian orbital elements allows to obtain \rfr{super} for $\dot\omega^\mathrm{2PN}$. It has a general validity, being straightforwardly applicable to whatsoever two-body system whose data are not analyzed within the DD framework, and clearly shows that the \textit{total} 2PN periastron precession does depend on the initial conditions, as confirmed also by the numerical integration of the 1PN+2PN equations of motion for a fictitious binary displayed in Figure\,\ref{fig1}. Also the formula for $k^\mathrm{1PN}+k^\mathrm{2PN}$ by \citet{1988NCimB.101..127D}, written in terms of the DD parameters, yields \rfr{super} if properly multiplied by the DD version of the 1PN mean motion and after appropriate conversion from the DD parameters to the osculating Keplerian ones.

For \psrab, $\dot\omega^\mathrm{2PN}$ is retrograde for any value of the initial orbital phase, as shown by the plot of \rfr{super} in Figure\,\ref{fig2}.
Since it turns out that $-0.00080^\circ\,\mathrm{yr}^{-1} \leq\dot\omega^\mathrm{2PN}\leq -0.00045^\circ\,\mathrm{\,yr}^{-1}$, it adds up to the spin--orbit Lense--Thirring precession $\dot\omega^\mathrm{LT,\,A} \simeq -0.0006^\circ\,\mathrm{yr}^{-1}$.

The value $\dot\omega^\mathrm{2PN} = 0.000439^\circ\,\mathrm{yr}^{-1}$ by \citet{2020MNRAS.497.3118H} comes from having neglected to multiply $k^\mathrm{1PN}$ by the 1PN mean motion and to expand the resulting product to the order of $\mathcal{O}\ton{c^{-4}}$.

For some astronomical systems in the Solar System of potential interest, the 2PN pericenter precession is negligible, while for the S-star S4711 orbiting Sgr A$^\ast$ it amounts to $-1.4\,\mathrm{''\,yr}^{-1}\leq\dot\omega^\mathrm{2PN}\leq  0.074\,\mathrm{''\,yr}^{-1}$.
\bibliography{psrbib}{}

\end{document}